\documentclass[12pt,preprint]{aastex}
\usepackage{xspace}
\usepackage{amsmath}
\usepackage{mathtools}
\usepackage{graphics,graphicx}
\newcommand{\as}{$^{\prime\prime}$\xspace}
\newcommand{\am}{$^{\prime}$\xspace}

\newcommand{\PL}{$\Gamma$\xspace}

\newcommand{\Cdof}{$C/d.o.f.$\xspace}
\newcommand{\Lx}{$L_{0.3-10}$\xspace}
\newcommand{\Msun}{$M_{\odot}$\xspace}

\newcommand{\abnd}{$Z/Z_{\odot}$\xspace}

\slugcomment{}
\shorttitle{The NGC~404 Central Engine}
\shortauthors{Binder et al.}

\begin{document}

\title{A Deep {\it Chandra} View of the NGC 404 Central Engine}
\author{B. Binder\altaffilmark{1}, B. F. Williams\altaffilmark{1}, M. Eracleous\altaffilmark{2}, A. C. Seth\altaffilmark{3}, J. J. Dalcanton\altaffilmark{1}, E. D. Skillman\altaffilmark{4}, D. R. Weisz\altaffilmark{1,}\altaffilmark{4}, S. F. Anderson\altaffilmark{1}, T. J. Gaetz\altaffilmark{3}, P. P. Plucinsky\altaffilmark{3}
\altaffiltext{1}{University of Washington, Department of Astronomy, Box 351580, Seattle, WA 98195}
\altaffiltext{2}{Department of Astronomy \& Astrophysics and Center for Gravitational Wave Physics, The Pennsylvania State University, 525 Davey Lab, University Park, PA 16802}
\altaffiltext{3}{Harvard-Smithsonian Center for Astrophysics, 60 Garden Street Cambridge, MA 02138, USA}
\altaffiltext{4}{University of Minnesota, Astronomy Department, 116 Church St. SE, Minneapolis, MN 55455}
}

\begin{abstract}
We present the results of a 100 ks {\it Chandra} observation of the NGC 404 nuclear region. The long exposure and excellent spatial resolution of {\it Chandra} has enabled us to critically examine the nuclear environment of NGC 404, which is known to host a nuclear star cluster and potentially an intermediate-mass black hole (on the order of a few times $10^5$ \Msun). We find two distinct X-ray sources: a hard, central point source coincident with the optical and radio centers of the galaxy, and a soft extended region that is coincident with areas of high H$\alpha$ emission and likely recent star formation. When we fit the 0.3-8 keV spectra of each region separately, we find the hard nuclear point source to be dominated by a power law (\PL = 1.88), while the soft off-nuclear region is best fit by a thermal plasma model ($kT$ = 0.67 keV). We therefore find evidence for both a power law component and hot gas in the nuclear region of NGC 404. We estimate the 2-10 keV luminosity to be 1.3$^{+0.8}_{-0.5}\times10^{37}$ erg s$^{-1}$. A low level of diffuse X-ray emission was detected out to $\sim$15\as ($\sim$0.2 kpc) from the nucleus. We compare our results to the observed relationships between power law photon index and Eddington ratio for both X-ray binaries and low luminosity active galaxies and find NGC 404 to be consistent with other low luminosity active galaxies. We therefore favor the conclusion that NGC 404 harbors an intermediate-mass black hole accreting at a very low level.
\end{abstract}

\section{Introduction}
It is well established that the nuclei of active galaxies exhibit a broad range of luminosities, from the most energetic quasars to more modest Seyferts. Active galactic nuclei (AGN) with nuclear X-ray luminosities below $\sim$10$^{41}$ erg s$^{-1}$ are classified as low luminosity AGNs (LLAGNs; Koratkar et al. 1995), and are the most common variety of AGN observed in the local universe. About 30\% of all nearby bright galaxies exhibit LLAGN activity, and many additionally host low-ionization nuclear emission-line regions (LINERs; Ho et al. 1997). Many recent studies have successfully linked LINERs to AGN activity; for example, the detections of X-ray cores (Dudik et al. 2005; Flohic et al. 2006; Gonz\'{a}lez-Mart\'{i}n et al. 2006; Zhang et al. 2009; Gonz\'{a}lez-Mart\'{i}n et al. 2009), radio cores (Nagar et al. 2005), and mid-IR coronal lines (Satyapal et al. 2004) are all indicative of an accreting black hole (BH) energy source. At low nuclear luminosities, however, several alternative explanations for the power source have been investigated, such as circumnuclear starbursts (Gonzales Delgado et al. 2004; Colina et al. 2002), shock heating by supernovae (SNe) in a high-density environment (Alonso-Herrero et al. 2000; Olsson et al. 2007), and photoionization by very hot O stars (Terashima et al. 2000). Moreover, a number of studies of the SEDs of weak AGNs in LINERs find that the AGN does not produce enough photons to power the emission lines (e.g., Ho 2008; Eracleous et al. 2010, and references therein).

At a distance of 3.1 Mpc (Karachentsev et al. 2004), NGC 404 is the nearest S0 type galaxy to the Milky Way and the closest galactic nucleus to be classified as a LINER by Ho et al. (1997). While the observed nuclear X-ray luminosity is low, only a few times 10$^{37}$ erg s$^{-1}$ (Eracleous et al. 2002), the presence of a LLAGN in NGC 404 remains ambiguous. The X-ray luminosity is consistent with that of a single high mass X-ray binary (HMXB) or a giant star-forming region such as 30 Doradus (Wang \& Helfand 1991), and the soft X-ray emission is consistent with a hot gas origin, potentially blown out by a compact starburst or SNe. No radio core has been observed at 15 GHz (to a limiting flux of 1.4 mJy; Nagar et al. 2005), however an unnresolved 3 mJy continuum source is detected at 1.4 GHz by del R\'{i}o et al. (2004), comparable in luminosity to the Crab Nebula. A compact X-ray source was previously detected in the central region of NGC 404 (Lira et al. 2000; Eracleous et al. 2002), but its low luminosity and soft thermal spectrum indicate a possible starburst event origin. Mid-IR observations of the NGC 404 nuclear region show high ionization lines consistent with AGNs (Satyapal et al. 2004); however, the [Ne V] lines (a more reliable indicator of AGN activity) are not detected (Abel \& Satyapal 2008).

Additionally, optical {\it Hubble Space Telescope} (HST) observations show H$\alpha$ emission occurring in both a compact source $0.^{\prime\prime}$16 north of the nucleus and in structures reminiscient of supernova remnants (Pogge et al. 2000), and [O III] emission originates from a double-lobed structure along the major axis of the galaxy (Plana et al. 1998) with a higher velocity dispersion than the central H$\alpha$ emission (Bouchard et al. 2010). While the UV spectrum of the nucleus reveals signatures of O stars, the dilution of the lines suggest that $\sim$60\% of the UV flux may originate from a nonthermal source (Maoz et al. 1998). The observed level of UV variability (the UV emission declined by a factor of 3 between 1993 and 2002; Maoz et al. 2005) provides the strongest evidence for the existence of an accreting, massive black hole in the NGC 404 nucleus.

Analysis of NICMOS data by Ravindranath et al. (2001) reveal a nuclear star cluster (NSC) within the central arcsecond of NGC 404. NSCs are present in $\sim$70\% of galaxies (Graham \& Guzman 2003; Mu\~{n}oz Mar\'{i}n et al. 2009), independent of the host galaxy morphology (B\"{o}ker et al. 2002; Carollo et al. 1997; C\^{o}t\'{e} et al. 2006), and are considered to be the foundation of circumnuclear starbursts and supernovae that could drive LINER activity (Meurer et al. 1995; Tremonti et al. 2001; Chandar et al. 2005). However, if NGC 404 is dominated by star formation, the rate is exceptionally low, with only two to six O stars being sufficient to explain the observed luminosity (del R\'{i}o et al. 2004). Dynamical modeling of stellar and gas kinematics in the nucleus by Seth et al. (2010) provide mixed evidence for the presence of a SMBH.  They derive a firm upper limit of $\sim10^6$ \Msun, and a best fitting gas dynamical mass of $4.5^{+3.0}_{-2.5}\times10^5$ \Msun (3$\sigma$ errors). Although other low-mass galaxies have been identified as candidate IMBH hosts through reverberation mapping (i.e., NGC 4395; Peterson et al. 2005) and indirect mass measurements (i.e., by narrow optical line measurements; see Greene \& Ho 2007), NGC 404 is potentially the lowest-mass central BH ever dynamically detected in the center of a galaxy. 

With both a NSC and possible IMBH, the nuclear region of NGC 404 is a complicated environment. However, NGC 404 provides an ideal test case to address several key questions relating to LLAGN activity: is there an intrinsic lower limit to the luminosity of the AGN phenomenon, and what fraction of LINERs are powered by stellar processes n versus those that host a dwarf version of more powerful Seyferts and quasars? Deep, high spatial resolution X-ray observations can potentially resolve many ambiguities surrounding the NGC 404 nucleus, such as the morphology of the X-ray emission, the shape of the X-ray spectrum, and the variability properties of the source.

In this paper, we present the analysis of a new, 100 ks {\it Chandra} observation of the NGC 404 nuclear region. In \S2, we provide a description of the observations and our data analysis procedures, including our imaging analysis, timing analysis, and spectral modeling. Our results are presented in \S3, and we give a discussion of our results in \S4. A summary of our work is given in \S5.

\section{Observations and Data Analysis}
NGC 404 was observed with {\it Chandra} ACIS-S on 2010 October 21-22 for a total useable time of 97 ks. The optical center of the NGC 404 nucleus centered on the S3 chip and detected at $\alpha_{\rm J2000}$ = 01$^{\rm h}$09$^{\rm m}$26.$^{\rm s}$99 and $\delta_{\rm J2000}$ = +35$^{\circ}$43\am05.$^{\prime\prime}$1 (with a positional uncertainty of $\sim$0.$^{\prime\prime}$02) in good agreement (within 0.$^{\prime\prime}$4) with the position reported in SIMBAD\footnote{See http://simbad.u-strasbg.fr/.} (Cotton et al. 1999) and  previous {\it Chandra} detections of the NGC 404 nucleus. The background count rate was estimated using an annular region devoid of any obvious sources, centered at the nucleus with an inner and outer radius of 25\as and 35\as, respectively. We find the 0.3-8 keV background count rate to be low, $\sim$5$\times10^{-6}$ ct s$^{-1}$ arcsec$^{-2}$, throughout the observation. We additionally obtained two archival {\it Chandra} ACIS-S observations of NGC 404 from 2000 Aug 30 (2 ks) and 1999 Dec 19-20 (24 ks). These observations are summarized in Table 1. 

All observations were reduced using the X-ray data analysis package CIAO version 4.3 and using standard reduction procedures. We created exposure maps for the images using the CIAO script \texttt{merge\_all}\footnote{See http://cxc.harvard.edu/ciao/ahelp/merge\_all.html.}. Point sources, including the NGC 404 nucleus, were identified using the CIAO task \texttt{wavdetect}\footnote{See http://cxc.harvard.edu/ciao/ahelp/wavdetect.html.}. We compared our X-ray point source detections to a 3.6$\mu$ {\it Spitzer} IRAC image of NGC 404 (aligned with the USNO-B1.0 star catalog) and find one X-ray source other than the NGC 404 nucleus to be coincident with a likely IR counterpart (with a positional offset of 0.$^{\prime\prime}$1), although this source is not a confirmed X-ray emitting object. 

We additionally searched for potential optical counterparts using optical observations in the Hubble Legacy Archive, but were unable to unambiguously identify any optical sources within the HST field of view coincident with our X-ray point sources. We thus conclude that our residual systematic uncertainty in absolute pointing is conservatively 0.$^{\prime\prime}$4.

All spectra were extracted using \texttt{psextract}\footnote{See http://cxc.harvard.edu/ciao/ahelp/psectract.html}, and spectral fitting was performed in \texttt{XSPEC} (Arnaud 1996) v.12.6.0q, and spectral models were fit to the unbinned spectra (binned spectra are shown for display purposes only). All models include a column of neutral absorption fixed at the Galactic value, $n_{\rm H,Gal}$ = $5.13\times10^{20}$ cm$^{-2}$ (Kalberla et al. 2005), estimated using the HEASARC $n_{\rm H}$ calculator.

We use $C$-statistics in lieu of traditional $\chi^2$ statistics, due to the low number of X-ray counts. Errors correspond to the 90\% confidence level. In addition to reporting the $C$-statistic and degrees of freedom (\Cdof) for each model, we used the \texttt{XSPEC} task {\it goodness} to perform Monte Carlo simulations of the spectra using each best-fit model. The procedure returns the percentage of the simulated spectra that had a fit statistic less than that obtained from the fit to the real data. A value of 50\% indicates the best-fitting model is a good representation of the data; values much less than this indicate that the data are overparameterized by the model (i.e., random statistical fluctuations in the majority of simulated spectra are not able to produce a fit statistic as low as that obtained from the real data), and values much higher than this indicate the model is a poor fit to the data (i.e., a large majority of simulated spectra have a fit statistic less than that obtained from the real data). We perform $5\times10^4$ realizations for each model, and we denote the resulting percentage as $MC$.

\section{Results}
\subsection{X-Ray Imaging and Hardness Ratio Maps}
We co-added the available {\it Chandra} observations to investigate the morphology of the NGC 404 X-ray emission. Earlier work by Eracleous et al. (2002) saw evidence for extended soft X-ray emission as far out as 10\as ($\sim$0.15 kpc) from the nucleus, with potential shell-like structures suggestive of a hot gas superbubble (see the discussion by Chu \& Mac Low 1990).

We divided our {\it Chandra} image into three energy bands: soft (0.3-1 keV), medium (1-2 keV), and hard (2-7 keV). Each image was adaptively smoothed using \texttt{csmooth}\footnote{See http://cxc.harvard.edu/ciao/ahelp/csmooth.html}. Figure 1 shows a smoothed RGB rendering of the NGC 404 nucleus. The image shows a hard core, coincident with the optical and radio center of the galaxy, and soft surrounding emission. Extended X-ray emission is detected out to $\sim$15\as ($\sim$0.2 kpc). We additionally see a soft source not detected in optical images, $\sim$10\as northeast of the nucleus. For comparison, the NSC extends out to 0.$^{\prime\prime}$7 ($\sim$10 pc; Seth et al. 2010), and the disk scale length of NGC 404 is $\sim$130\as ($\sim$2 kpc; Baggett et al. 1998). The adaptively smoothed 0.3-1 keV and 2-7 keV images, with contours superimposed, are shown in Figure 2 to emphasize the compact nature of the hard central source.


We next used the smoothed soft and hard energy images to construct a hardness ratio map of the NGC 404 nuclear region (Figure 3). We define a hardness ratio HR = (hard-soft)/(hard+soft); dark areas in the hardness ratio map correspond to softer X-ray emission, while light regions indicate hard emission. There is clearly a point-like region of hard emission, coincident with the location of the NSC, surrounded by an extended area of predominantly soft X-ray emission.

In Figure 4, we show a H$\alpha$-$I$ color map of the NGC 404 nuclear region (Seth et al. 2010), with contours from our HR map overlaid. Dark regions on the H$\alpha$-$I$ color map correspond to regions with high H$\alpha$ flux, likely associated with young stars. We find one such region is coincident with soft, extended X-ray emission seen in our RGB rendering.

\subsection{Timing}
We find no evidence for long-term ($\sim$10 year) variability; our best-fit 0.5-2 keV luminosity ($\sim$10$^{37}$ erg s$^{-1}$, see next section) agree with that found by Eracleous et al. (2002). We generated light curves in four energy ranges (the total 0.3-10 keV band, and the soft, medium, and hard bands described above) to search for short-term variability over the course of our 97 ks observation. We generated cumulative arrival time distributions of counts for each light curve, then ran a two-sided KS test against the expected cumulative arrival time distrbution for a constant count rate. 

Figure 5 shows the cumulative arrival time distributions for our total 0.3-8 keV light curve and the hard 2-7 keV light curve. We ran a two-sided KS test for each of our cumulative arrival time distributions against a constant count rate. As summarized in Table 2, we found no evidence from our KS tests for variability at energies softer than 2 keV; however, our hard light curve yields a KS chance probability of 5.8$\times10^{-4}$. Both AGNs and XRBs exhibit strong and rapid variability; the apparent  detection of hard X-ray variability on time scales of $\sim$1 day therefore lends support to the idea that the NGC 404 nucleus hosts an accreting object. Detailed spectral fitting (described in the next section) is needed to constrain the nature of the compact accreting object further.

\subsection{X-ray Spectral Fitting}
Hard X-ray spectra of LLAGNs in LINERs are typically well represented by a two-component model: a power-law component plus soft thermal emission (Terashima et al. 2000), and the H$\alpha$ luminosoties of LINERs are positively correlated with the X-ray luminosities in the 2-10 keV band (Ho 2001). To test whether the NGC 404 nucleus is consistent with being a LLAGN, we use our observation with archival {\it Chandra} ACIS-S observations to perform spectral fitting in the 0.3-8 keV energy band. The total nuclear region ($<$17\as) of NGC 404 contained $\sim$1200 counts.

Our imaging analysis of the NGC 404 nuclear region indicated the presence of two distinct regions within the nucleus: a hard point source coincident with the optical and radio center of the galaxy, and a soft, extended region likely associated with an area of recent star formation. Using our HR contours and H$\alpha$-$I$ color map, we extracted spectra using elliptical regions for each of the two distinct sources, with an elliptical area of 11 square arcseconds for the hard point source and 27 square arcseconds for the soft extended source. The hard nuclear point source contained $\sim$500 counts, and our soft, extended region contained $\sim$90 counts. 

All our models fix an absorption component due to the Galactic column and an intrinsic absorption inferred from optical extinction from Schlegel et al. (1998). We find no evidence in our spectral fitting for additional absorption, and we find no evidence for the presence of a reflection component in any of the 0.3-8 keV spectra. We use the mass-metallicity relation derived in Tremonti et al. (2004) to estimate the NGC 404 metallicity to be 12 + log(O/H) $\sim$8.6-9 (i.e., near solar abundances).

We first attempted to model the soft, diffuse emission as a power law, a disk blackbody, and a thermal plasma (\texttt{APEC}; Smith et al. 2001). We find the thermal plasma model provides the best fit, with \Cdof = 5.5/4 and $MC$ = 51\% for $kT$ = 0.67$\pm0.11$ keV and abundances fixed at their solar values. The 0.3-10 keV luminosity of the soft X-ray emission is (2.5$\pm0.2)\times10^{36}$ erg s$^{-1}$, with an estimated 2-10 keV luminosity of 8.8$^{+0.4}_{-3.8}\times$10$^{34}$ erg s$^{-1}$. Our data are consistent with the idea of gas being ejected from the central region in a superbubble.

We attempted to model the hard, nuclear point-source as a simple power law, but were unable to obtain an acceptable fit (\Cdof = 33/25 and $MC$ = 76\%). Single-temperature thermal plasma models, with abundances either fixed at their solar values or allowed to vary, additionally did not result in acceptable fits (\Cdof = 68/25 for \abnd fixed at solar, and \Cdof = 42/24 with \abnd$<$0.05). Two-temperature thermal plasma models, with abundances fixed at solar values or allowed to vary, severely over-parameterized the data. We find $MC$ = 0.4\% with \Cdof = 11/23 for our model with \abnd = 1, and $MC$ = 0.3\% with \Cdof = 11/22 for \abnd$<$0.06.

We next attempted to model the hard, nuclear point source as a power law contaminated by thermal emission from the soft, diffuse X-ray source. For each fit, we fixed the thermal plasma temperature to be within the 90\% confidence interval of our best-fit model to the soft, diffuse source. Abundances were kept fixed at solar values. We find the hard, nuclear point source to be best described by $kT$ = 0.78 keV, with a power law photon index \PL = 1.88$^{+0.28}_{-0.32}$ contributing to $\sim$72\% of the total 0.3-8 keV photon flux and \Cdof = 21/24. We find a 0.3-10 keV luminosity of 2.4$^{+0.7}_{-0.4}\times10^{37}$ erg s$^{-1}$ and a 2-10 keV luminosity of 1.2$^{+0.7}_{-0.4}\times10^{37}$ erg s$^{-1}$. Although the fit moderately overparameterizes the data ($MC$=21\%), this model produces our best fitting parameters and is consistent with our imaging analysis.

Finally, we applied the results of our spectral fitting to the soft, diffuse emission and the hard, nuclear point source to model the entire nuclear region of NGC 404 as a power law contaminated by thermal plasma emission. We assume the best-fit thermal plasma temperature (0.67 keV) from the soft, diffuse emission and the best-fit power law photon index (\PL = 1.88) from the hard nuclear point source. We find the power law component contributes $\sim$55\% of the 0.3-8 keV photon flux, with \Cdof = 124/126, with $MC\sim$30\%. To estimate the errors on the power law photon index, we find the best-fit value of \PL when the thermal plasma temperature is set at our 90\% confidence interval lower limit (0.56 keV) and upper limit (0.78 keV). We find a best-fit photon index \PL = 1.85$^{+0.31}_{-0.30}$ using this approach. We estimate the 0.3-10 keV luminosity to be 3.0$^{+0.7}_{-0.5}\times10^{37}$ erg s$^{-1}$ and a 2-10 keV luminosity of 1.3$^{+0.8}_{-0.5}\times10^{37}$ erg s$^{-1}$.

The results of our spectral fitting are summarized in Table 3. Figure 6 shows the 0.3-8 keV spectra for the soft, diffuse emission, the hard nuclear point source, and the total nuclear region (with our best-fit models superimposed).

The presence of a power law component in the NGC 404 nucleus, in addition to variability in the 2-7 keV emission, lends support to the idea that the NGC 404 nucleus hosts an accreting black hole, but its low luminosity, on the order of a few times 10$^{37}$ erg s$^{-1}$, is comparable to that of a single XRB.

\section{Discussion}
By modeling the hard nuclear point source and diffuse soft emission found in the NGC 404 nucleus separately, we are able to resolve the ambiguity of the X-ray emission: a hard point source provides a power law component, and extended, diffuse gas supplies the thermal plasma emission. 

The Eddington luminosity is defined as $L_{\rm Edd}$ = 1.3$\times10^{38}$ ($M_{\rm BH}$/\Msun) erg s$^{-1}$, and the Eddington ratio $\xi$ is commonly defined as $\xi$ = log$_{10}$($L/L_{\rm Edd}$), where the bolometric luminosity $L$ is typically estimated as $L/L_{0.2-25 \rm keV}$ = 16 for AGNs (Ho 2008), whereas the bolometric correction factor for XRBs is roughly 2-5 times lower than for AGN (Wu \& Gu 2008; hereafter, WG08). The relationship between the X-ray power law photon index \PL and Eddington ratio has been investigated for XRBs (WG08) and low luminosity AGNs (LLAGNs; Constantin et al. 2009; hereafter C+09). An anticorrelation is found for LLAGNs, whereas a positive correlation is observed for XRBs and luminous AGNs (Wang et al. 2004; Shemmer et al. 2006).

We estimate $\xi$ for NGC 404 assuming the black hole is an XRB, with $M_{\rm BH}\sim$10 \Msun, and an IMBH AGN, with $M_{\rm BH}\sim10^5$ \Msun. In Figure 7, we use our best-fit photon index for the hard nuclear point source and estimates of $\xi$ to compare our NGC 404 data to the observed relationships for both XRBs and LLAGNs from WG08 and C+09, respectively. The errors in $\xi$ indicate a factor of 3 change in BH mass (i.e., an AGN ranging from 3.3$\times10^4$ \Msun to 3$\times10^5$ \Msun and an XRB ranging from 3.3 \Msun to 30 \Msun). We find that while our data fall well below the WG08 relationship for a high/soft state XRB, our data are consistent with the anticorrelation found for the C+09 LLAGN sample. However, due to the large scatter observed in the LLAGN sample and the errors in our observed photon index, we cannot decisively rule out the possibility that the NGC 404 central engine is powered by an XRB in the low/hard state. Our data moderately favor the IMBH AGN interpretation. Although the current work moderately favors the low-mass AGN interpretation, the UV spectrum and nuclear star cluster still allow the possibility of an XRB component.

In Table 4, we summarize the observed multiwavelength properties of NGC 404, found in both the literature and presented in this work, and indicate if the origin is likely to be an AGN or XRB. Additionally, the upper limits on the radio core (Nagar et al. 2005) and the detection of an unresolved radio continuum (del R\'{i}o et al. 2004) can be combined with our deep X-ray observations to place upper limits on the central BH mass of NGC 404 -- a correlation has been established relating the radio luminosity $L_R$ and X-ray luminosity $L_X$ (Corbel et al. 2003; Gallo, Fender \& Pooley 2003; Gallo et al. 2006) cover many orders of magnitude in BH mass and luminosity, forming the ``fundamental plane of black hole activity.'' We use the best-fit BH ``fundamental plane'' recently presented by Bell et al. (2011) and the upper limit on the radio core flux of NGC 404 to estimate $M_{\rm BH}<2\times10^6$ \Msun. If the unresolved 3 mJy continuum source is indeed powered by an accreting BH, it would imply $M_{\rm BH}\sim3\times10^5$ \Msun.

Additionally, we use the following relation between radio luminosity and star formation rate (Condon 1992), 

\begin{equation}
\left(\frac{L_{\rm N}}{\text{W Hz$^{-1}$}}\right)\sim5.3\times10^{21} \left(\frac{\nu}{\text{GHz}}\right)^{-\alpha} \left[\frac{SFR(M\geq5 M_{\odot})}{M_{\odot} \text{yr}^{-1}}\right],
\end{equation}

\noindent where $\alpha\sim0.8$ is the nonthermal spectral index and $SFR$ is the star formation rate (in \Msun yr$^{-1}$), to test whether the observed radio luminosity is consistent with the observed low star formation rate. Using the $SFR$ upper limit ($\sim10^{-3}$ \Msun yr$^{-1}$) estimated in Seth et al. (2010), we predict an upper limit on the radio luminosity at 1.4 GHz to be $\sim$4$\times10^{25}$ erg s$^{-1}$. This upper limit is roughly eight orders of magnitude below the observed 1.4 GHz upper limit for the NGC 404 nucleus. We therefore conclude the observed radio luminosity cannot be explained by star formation alone, and is evidence for the presence of an AGN in the NGC 404 nucleus.

The X-ray luminosity of the AGN in NGC 404 bears directly on the question of whether accretion power can account for the observed luminosities of the optical emission lines, whose relative intensities are the defining characteristic of LINERs. In a recent study of the energy budgets of three dozen LINERs, including NGC 404, Eracleous et al. (2010) found that in the majority of cases the weak AGN does not provide enough ionizing photons to account for the observed luminosities of the hydrogen recombination lines. This conclusion is in general agreement with previous studies, as discussed in Eracleous et al. (2010). In the particular case of NGC 404 the number of ionizing photons was found to be deficient by a factor of $\approx 60$. The X-ray luminosity of NGC 404 measured here is only $\approx 25$\% higher than that measured by Eracleous et al. (2002), after accounting for the different distance used in that paper, and is consistent (within errors) with the result obtained here. Therefore, the situation regarding the ionizing photon output of the AGN remains the same. However, Maoz et al. (1998) have estimated the ionizing luminosity of hot stars in the nucleus of NGC 404 based on their measurements of the UV spectrum with the {\it HST} and found it to be adequate to power the emission lines. This conclusion was re-iterated by Seth et al. (2010), who also noted that this ionizing luminosity could be provided by a relatively small number of O stars in the nuclear star cluster. Therefore, the LINER in NGC 404 appears to be powered by stellar processes.

\section{Summary}
We present the results of a 100 ks {\it Chandra} observation of the nearby LINER and S0 galaxy NGC 404. The deep exposure has allowed us to critically test several forms for the 0.3-10 keV spectrum, and the excellent spatial resolution of {\it Chandra} has enabled us to investigate the X-ray morphology of the NGC 404 nuclear region. We find the 0.3-10 keV spectrum to be consistent with emission from hot gas plus a power law continuum, and we are specifically able to separate a point source of high energy photons from a diffuse source of soft X-ray emission. Additionally, find evidence for variability in the hard 2-10 keV emission.

The presence of a power law component and a moderate level of variability in the hard emission is indicative of X-ray emission powered by accretion onto a BH. The estimated 0.3-10 keV luminosity ($\sim$2-3$\times10^{37}$ erg s$^{-1}$) is both comparable to that of a single Galactic XRB and consistent with a IMBH accreting at extremely low levels, on the order of a few times 10$^{-9}$ \Msun yr$^{-1}$. We estimate the Eddington ratio for both scenarios (assuming a 10 \Msun XRB and a 10$^5$ \Msun AGN) and compare our best-fit photon index \PL of the hard nuclear point source to the observed trends with $\xi$ for both XRBs and LLAGN. We find the NGC 404 X-ray spectral shape and luminosity to be consistent with observed LLAGNs, and inconsistent with observed XRBs. We therefore favor the scenario in which the NGC 404 nucleus is powered by an IMBH, with a mass on the order of $10^5$ \Msun as dynamically estimated by Seth et al. (2010). Such a weak AGN does not produce a sufficient quantity of ionizing photons necessary to power a LINER -- we therefore conclude that the LINER in NGC 404 is powered by stellar processes.

Very low accretion rates are common in nearby galaxies with BH masses less than a few times 10$^6$ \Msun (e.g., Baganoff et al. 2001; Garcia et al. 2000; Ho et al. 2003). Additional, multiwavelength observations of the NGC 404 nucleus are required to robustly determine the mass of the cental BH -- for example, resolved stellar populations within the nucleus would enable a robust dynamical mass determination, and a radio detection of the compact source would verify the location of NGC 404 on the fundamental plane of BH activity.

\acknowledgements
The authors would like to thank the anonymous referee for helpful comments. B. Binder and B. F. Williams acknowledge support from {\it Chandra} grant GO1-12118X. T. Gaetz and P. P. Plucinsky acknowledge support of NASA Contract NAS8-03060.


\begin{table}[!ht]\footnotesize
\caption{{\it Chandra} ACIS-S Observations}
\centering
\begin{tabular}{ccccc}
\hline\hline
Date & ObsID & R.A. & Decl. & Exposure \\
     &       & (J2000) & (J2000) & (ks) \\
(1)  & (2)   & (3)  & (4)    &  (5)      \\
\hline 
2010 Oct 22  & 12239 & 01 09 26.99 & +35 43 05.1 & 97 \\
2000 Aug 30  & 384   & 01 09 27.06 & +35 43 05.0 & 2 \\
1999 Dec 19-20 & 870 & 01 09 26.94 & +35 43 05.5 & 24 \\
\hline\hline
\end{tabular}
\end{table}

\begin{table}[!ht]\footnotesize
\caption{Temporal Variability as a Function of Energy}
\centering
\begin{tabular}{ccc}
\hline\hline
Energy range (keV) & Net Counts & K-S chance probability \\
(1)                & (2)                 & (3)\\
\hline 
0.3-10  & 1237 & 0.756 \\
\hline
0.3-1   & 487 & 0.811 \\
1-2     & 261 & 0.365\\
2-7     & 220 & 5.8$\times10^{-4}$ \\
\hline\hline
\end{tabular}
\end{table}


\begin{table}[!ht]\footnotesize
\caption{Best-Fit Spectral Models for the NGC 404 0.3-8 keV Nucleus}
\centering
\begin{tabular}{ccccc}
\hline\hline
Region & Best-Fit Model & Parameter & Best-Fit Value \\
(1)    & (2)            & (3)       & (4)            \\
\hline
Soft, diffuse & thermal & $kT$  & 0.67$\pm$0.11 keV \\
emission      & plasma  & \Cdof & 5.5/4 \\
              &         & $MC$  & 51\% \\
              &         & \Lx   & (2.5$\pm0.2)\times$10$^{36}$ erg s$^{-1}$ \\
              &         & $L_{2-10}$ & 8.8$^{+0.4}_{-3.8}\times$10$^{34}$ erg s$^{-1}$ \\
\hline
Hard, nuclear & power law +    & $kT$     & 0.78 keV (fixed) \\
point source  & thermal plasma & \PL      & 1.88$^{+0.28}_{-0.32}$ \\
              &                & frac. PL & 72\% \\
              &                & \Cdof    & 21/24 \\
              &                & $MC$     & 21\% \\
              &                & \Lx      & 2.4$^{+0.7}_{-0.4}\times10^{37}$ erg s$^{-1}$ \\
              &               & $L_{2-10}$ & 1.2$^{+0.7}_{-0.4}\times10^{37}$ erg s$^{-1}$ \\
\hline
Total nuclear & power law +    & $kT$     & 0.67 keV (fixed) \\
region        & thermal plasma & \PL      & 1.85$^{+0.31}_{-0.30}$ \\
              &                & frac. PL & 55\% \\
              &                & \Cdof    & 124/126 \\
              &                & $MC$     & 30\% \\
              &                & \Lx      & 3.0$^{+0.7}_{-0.5}\times10^{37}$ erg s$^{-1}$ \\
              &               & $L_{2-10}$ & 1.3$^{+0.8}_{-0.5}\times10^{37}$ erg s$^{-1}$ \\
\hline\hline
\end{tabular}
\end{table}


\begin{table}[!ht]\footnotesize
\caption{AGN vs. XRB Properties Exhibited by the NGC 404 Nucleus}
\centering
\begin{tabular}{ccc}
\hline\hline
Property or Observation & AGN & XRB \\
(1)  & (2)   & (3)   \\
\hline
Radio fluxes and upper limits & \checkmark & \\
Soft X-ray emission$^{\dag}$ &  & \checkmark \\
Supernova remnant-like optical H$\alpha$ emission$^{\dag}$ & & \checkmark \\
Mid-IR high ionization lines; hot dust & \checkmark & \\
UV spectrum & \checkmark & \checkmark \\
UV variability & \checkmark & \\
Dynamical BH estimates & \checkmark & \\
Nuclear star cluster & & \checkmark \\
X-ray 2-10 keV variability; {\it this work} & \checkmark & \checkmark \\
X-ray power law emission (\PL = 1.88$^{+0.28}_{-0.32}$); {\it this work} & \checkmark &  \\
\hline\hline
\multicolumn{3}{l}{$^{\dag}$Observations are {\it not} associated with the 
hard nuclear point source.} \\
\end{tabular}
\end{table}


\begin{figure}[!ht]
\centering
\includegraphics[width=0.7\linewidth,clip=]{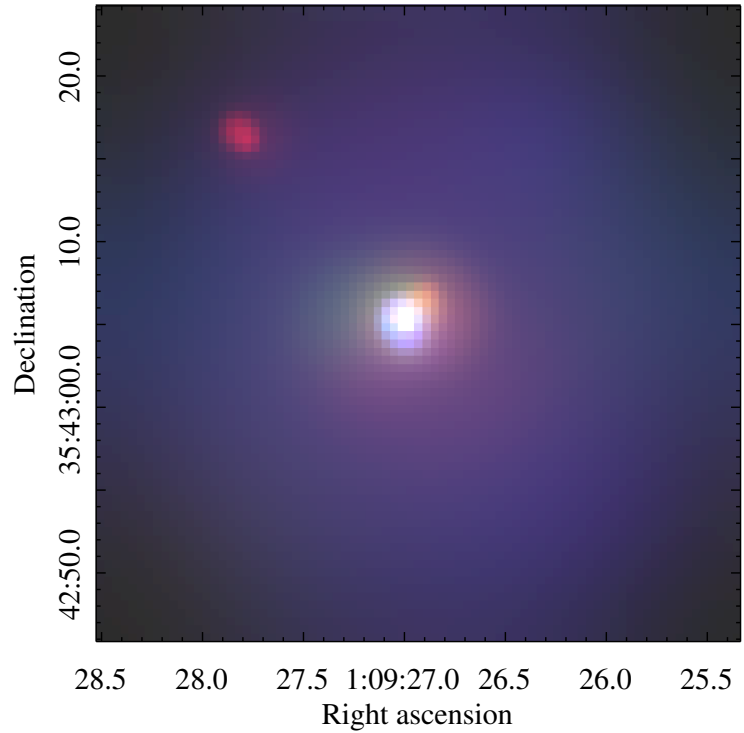}
\caption{Adaptively smoothed image of our {\it Chandra} observation of 
the NGC 404 nuclear region. This rendering emphasizes the hard nuclear point 
source and the soft extended region in the center of the galaxy, as well 
as a super-soft X-ray source $\sim$10\as to the northeast of the nucleus 
and low levels of diffuse emission out to $\sim$15\as. Red = 0.3-1 keV, 
green = 1-2 keV, and blue = 2-7 keV.}
\end{figure}


\begin{figure}[!ht]
\centering
\begin{tabular}{cc}
\includegraphics[width=0.5\linewidth]{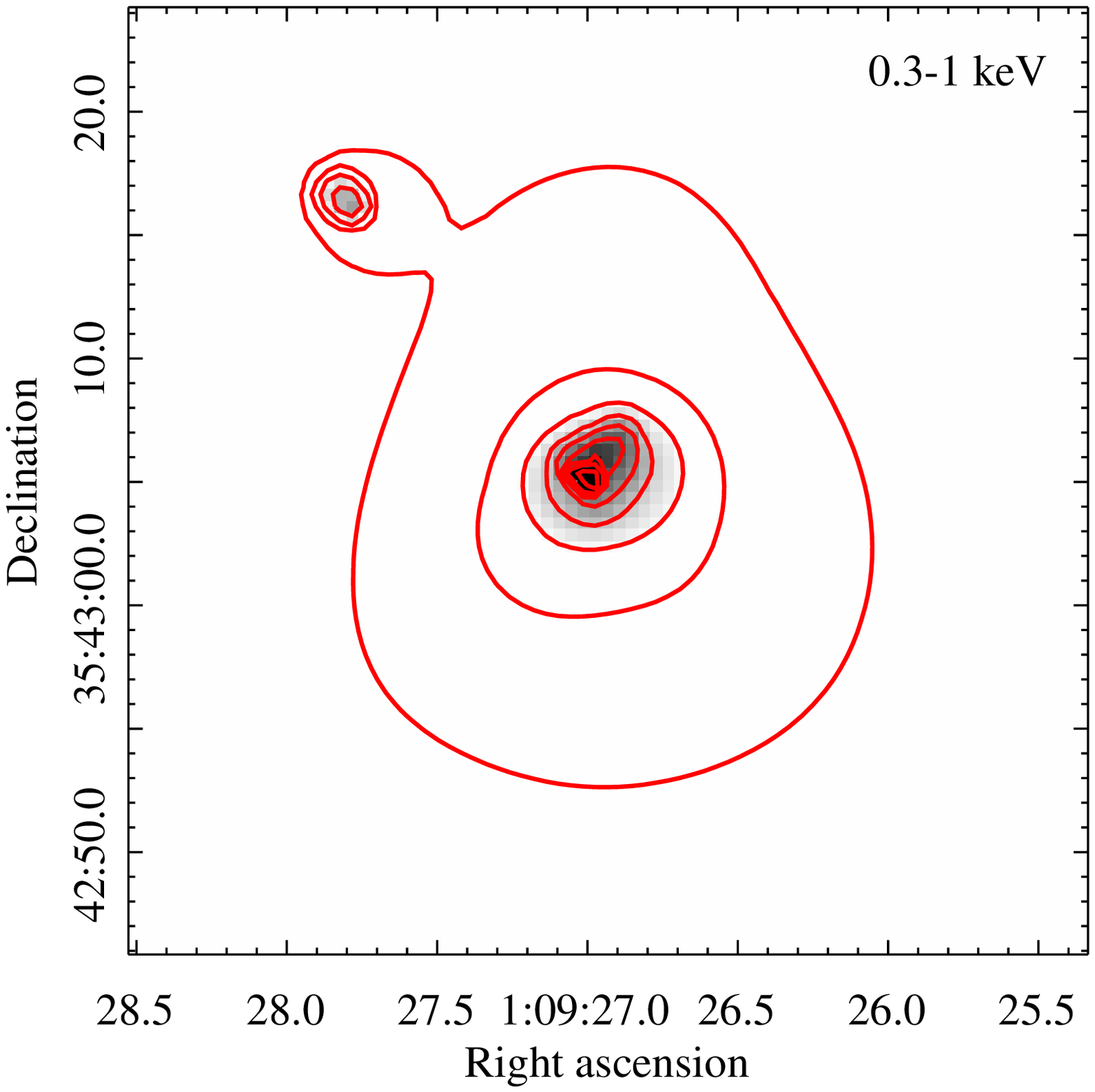} &
\includegraphics[width=0.5\linewidth]{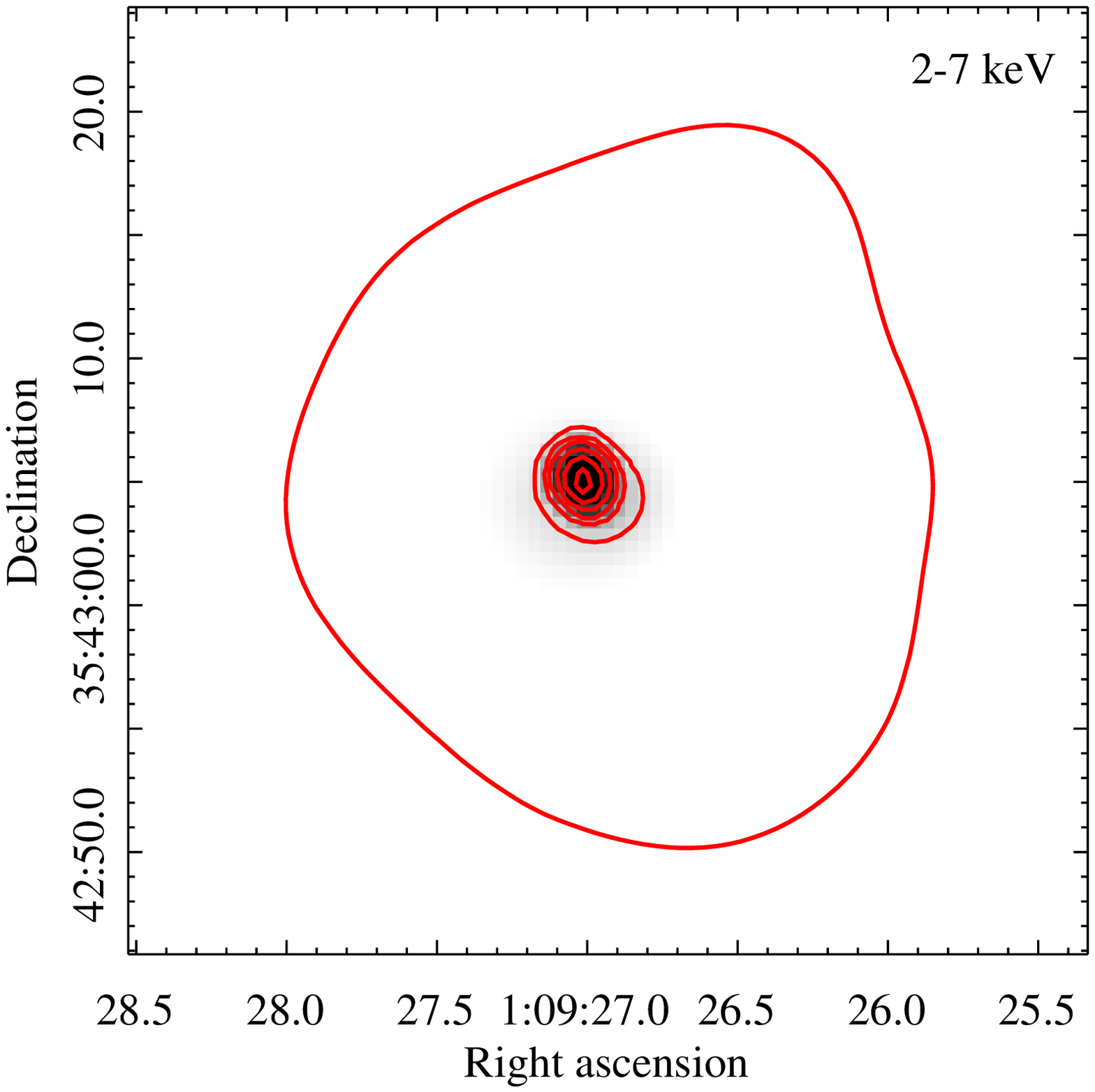} \\
\end{tabular}
\caption{The {\it left} panel shows the smoothed 0.3-1 keV emission, and 
the {\it right} panel shows the smoothed 2-7 keV emission. The red countours 
are set to the same levels for both images. The soft, 0.3-1 keV contours 
show extended, asymmetric emission, while the hard 2-7 keV contours are 
consistent with a point source origin.}
\end{figure}


\begin{figure}[!ht]
\centering
\includegraphics[width=0.7\linewidth,clip=]{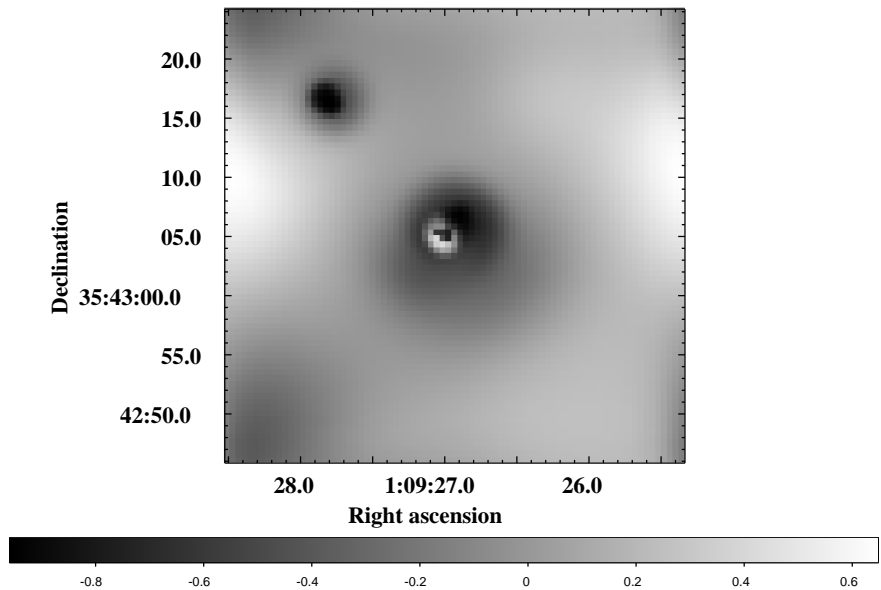} \\
\caption{The smoothed hardness map of the NGC404 nuclear X-ray emission. Dark 
regions indicate soft emission, while light regions indicate hard emission. A 
hard point source is clearly visible within a region of extended soft 
emission.}
\end{figure}


\begin{figure}[!ht]
\centering
\includegraphics[width=0.7\linewidth,clip=]{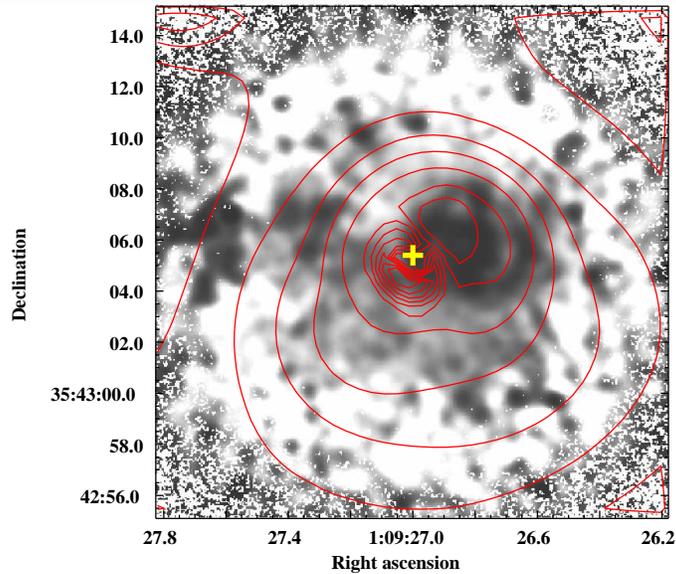} 
\caption{The HST H$\alpha$-$I$ color map of the NGC 404 nucleus with HR map 
contours superimposed (red). Dark regions in the map are likely associated 
with areas of younger stars, and appear to coincide with the soft, extended 
X-ray emission region seen in our HR map and RGB rendering. The center of 
the radio emission is shown by the yellow cross, and coincides with the 
center of the hard nuclear point source.}
\end{figure}


\begin{figure}[!ht]
\centering
\begin{tabular}{cc}
\includegraphics[width=0.43\linewidth,clip=]{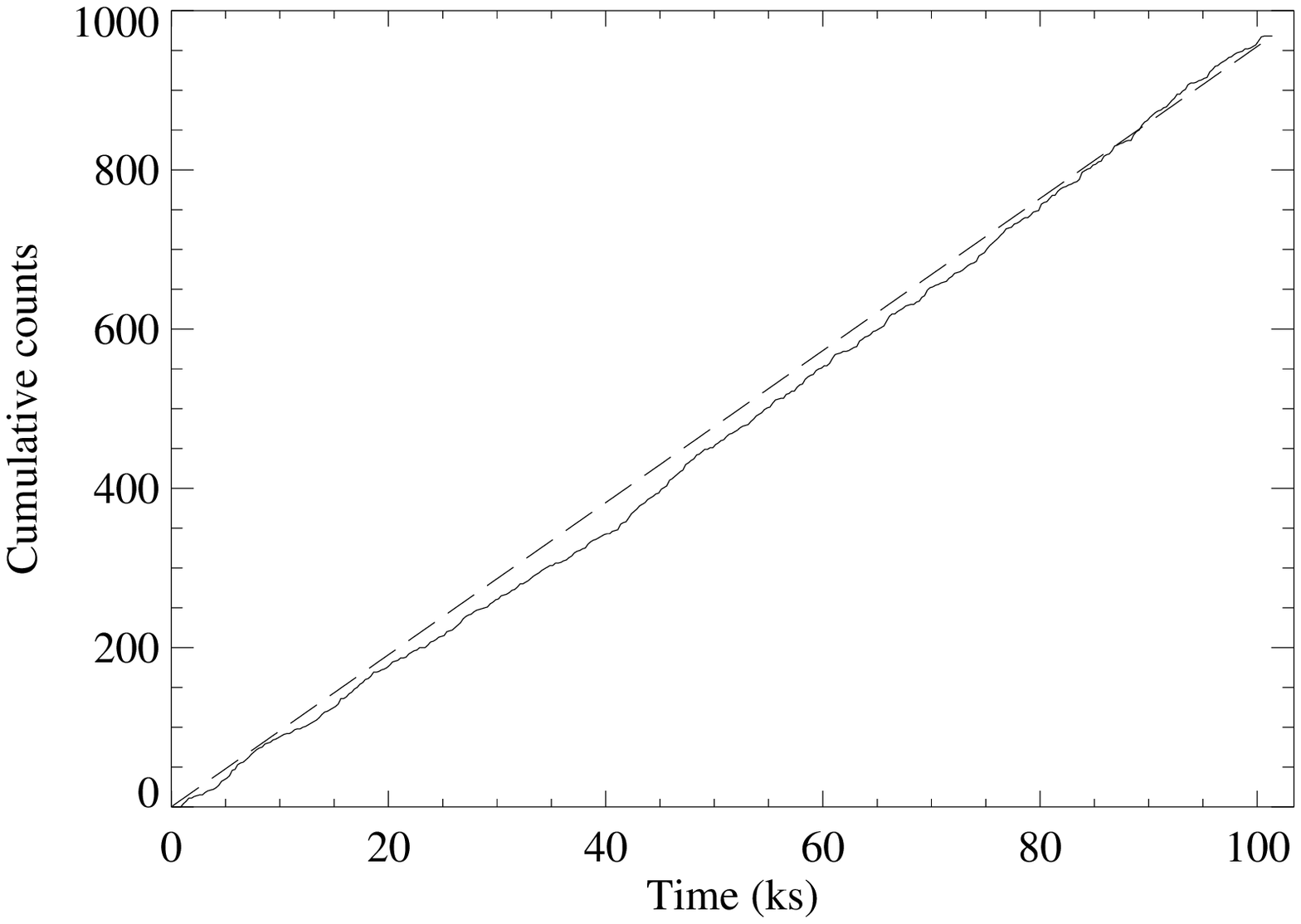} &
\includegraphics[width=0.43\linewidth,clip=]{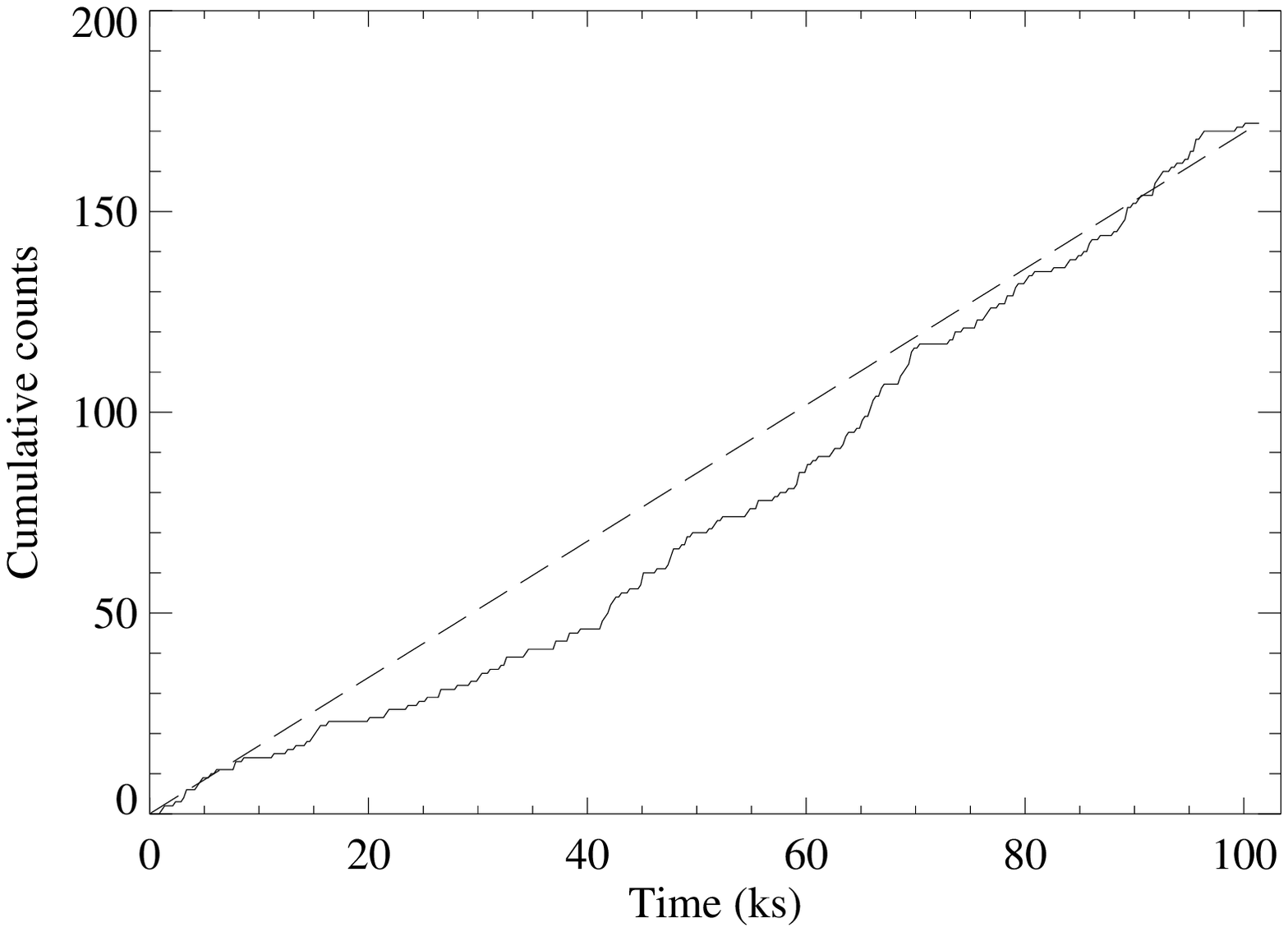} \\
\end{tabular}
\caption{Cumulative arrival time of counts during the observation. The dashed 
lines represent the predicted arrival times assuming a constant count rate. 
The {\it left} panel is for the total 0.3-8 keV light curve, and is 
consistent with a constant count rate. The {\it right} panel is for the 
hard 2-7 keV light curve, and shows evidence for variability.}
\end{figure}


\begin{figure}[!ht]
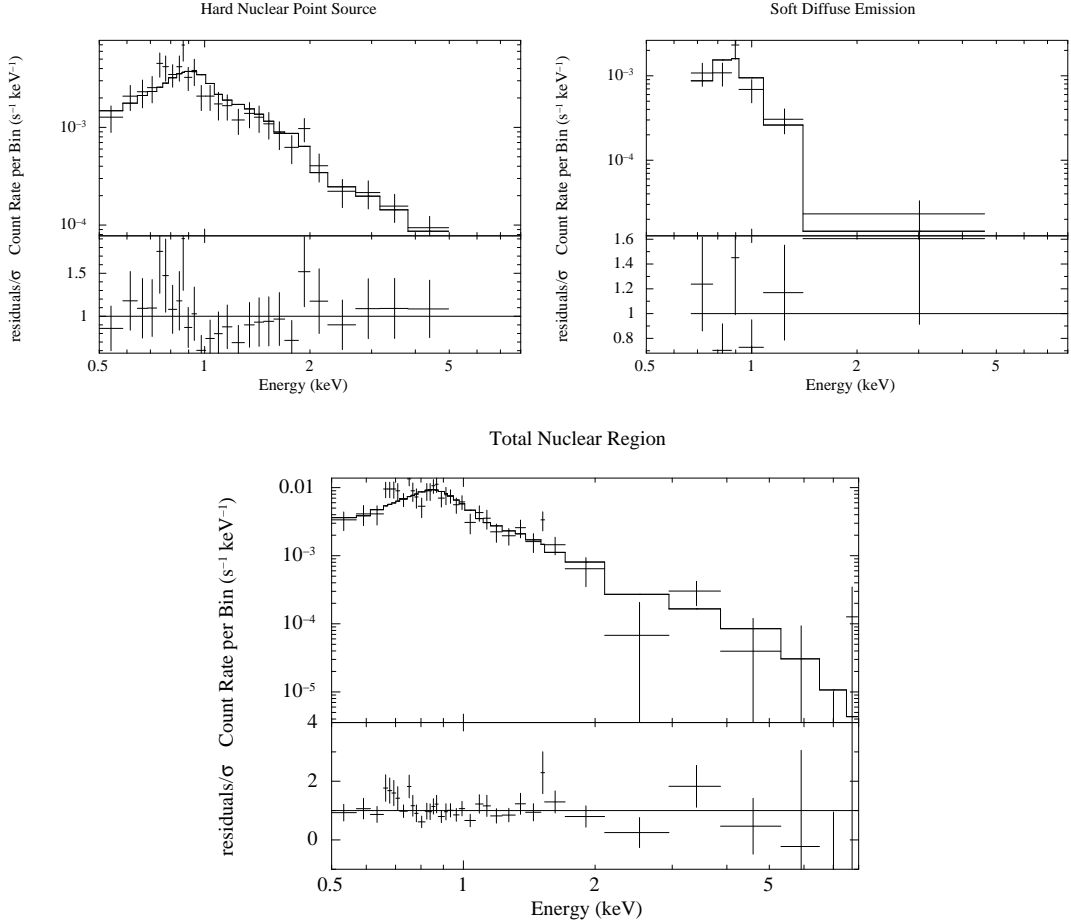

\centering
\begin{tabular}{cc}
\includegraphics[width=0.32\linewidth,clip=,angle=-90]{fig6a.ps} &
\includegraphics[width=0.32\linewidth,clip=,angle=-90]{fig6b.ps} \\
\multicolumn{2}{c}{\includegraphics[width=0.4\linewidth,clip=,angle=-90]{fig6c.ps}} \\
\end{tabular}
\caption{{\it Top left}: The 0.5-8 keV hard nuclear point source spectrum, 
containing $\sim$500 counts, with the best-fit power law and thermal plasma 
model superimposed. {\it Top right}: The 0.5-8 keV soft, off-nuclear source 
spectrum, containing $\sim$90 counts, with the best-fit thermal plasma 
model superimposed. {\it Bottom}: Thetotal NGC 404 nucleus, with our best-fit 
model (a thermal plasma diluted by a power law) superimposed. Each model was 
fit to the total, unbinned spectrum -- the spectra have been binned here for 
display purposes only.}
\end{figure}


\begin{figure}[!ht]
\centering
\includegraphics[width=0.65\linewidth,clip=]{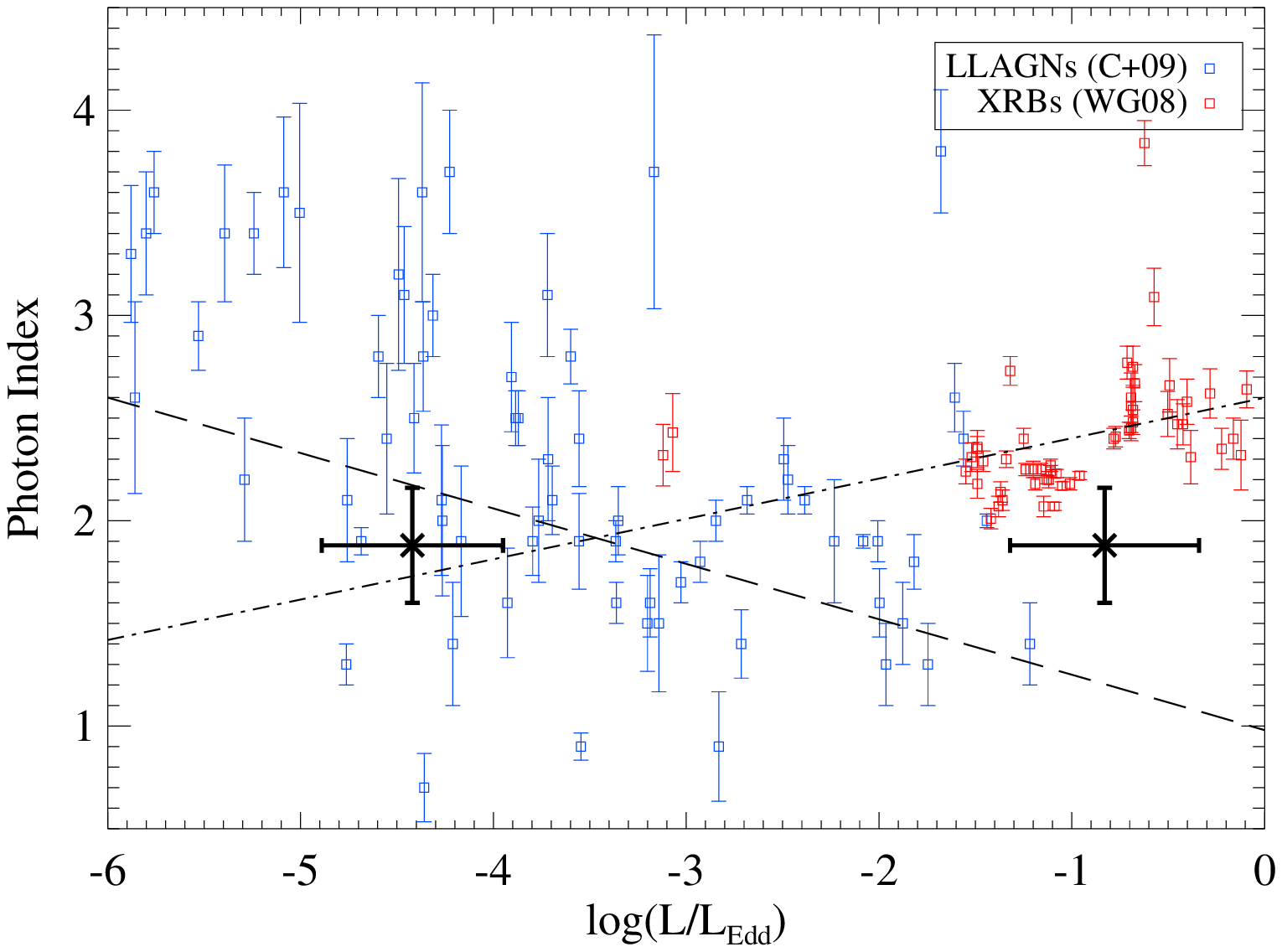}
\caption{The observed relationship between power law photon index and 
Eddington ratio for LLAGNs (C+09, blue points) and XRBs (WG08, red points). 
The dashed line shows the best-fitting anticorrelation to the 
LLAGN sample, and the dot-dashed line shows the best-fitting correlation 
for the XRB sample. The estimated Eddington ratio for NGC 404, assuming a 
10 \Msun and 10$^5$ \Msun central black hole, along with the best-fit 
photon index from our spectral fitting, is shown by the black crosses. 
Assuming a low Eddington ratio places NGC 404 well within the typical 
values for LLAGNs, while assuming a high Eddington ratio places NGC 404 
in a vacant part of the diagram. Our data are consistent with an IMBH 
central engine powering the X-ray emission from the NGC 404 nucleus.}
\end{figure}


\begin{references}
\reference{Ab08} Abel, N. P. \& Satyapal, S. 2008, ApJ, 678, 686
\reference{AH00} Alonso-Herrero, A., Rieke, M. J., Rieke, G. H., \& Shields, 
J. C. 2000, ApJ, 530, 688
\reference{Ar96} Arnaud, K. A. 1996 ASPC, 101, 17
\reference{Ba01} Baganoff, F. K. et al. 2001, Nature, 413, 45
\reference{Ba98} Baggett, W. E., Baggett, S. M., \& Anderson, K. S. J. 1998, 
AJ, 116, 1626
\reference{Be11} Bell, M. E. et al. 2011, MNRAS, 411, 402
\reference{Bo02} B\"{o}ker, T., Laine, S., van der Marel, R. P., Sarzi, M., 
Rix, H.-W., Ho, L. C., \& Shields, J. C. 2002, AJ, 123, 1389
\reference{Bo10} Bouchard, A., Prugniel, Ph., Koleva, M., \& Sharina, M. 
2010, A\&A, 513, 54
\reference{Ca97} Carollo, C. M., Stiavelli, M., de Zeeuw, P. T., \& Mack, J. 
1997, AJ, 114, 2366
\reference{Ch05} Chandar, R., Leitherer, C., Tremonti, C. A., Calzetti, D., 
Aloisi, A., Meurer, G. R., \& de Mello, D. 2005 ApJ, 628, 210
\reference{Ch90} Chu, Y.-H. \& Mac Low, M.-M. 1990, ApJ, 365, 510
\reference{Co02} Colina, L., Gonz\'{a}lez Delgado, R., Mas-Hesse, J. M., 
Leitherer, C. 2002, ApJ, 579, 545
\reference{Co92} Condon, J. 1992, ARAA, 30, 575
\reference{Co09} Constantin, A., Green, P., Aldcroft, T., Kim, D.-W., 
Haggard, D., Barkhouse, W., \& Anderson, S. F. 2009, ApJ, 705, 1336
\reference{Co99} Cotton, W. D., Cotton, J. J., \& Arbizzani, E. 1999, 
ApJS, 125, 409
\reference{Co03} Corbel, S., Nowak, M. A., Fender, R. P., Tzioumis, A. K., \& 
Markoff, S. 2003, A\&A, 400, 1007
\reference{Co06} C\^{o}t\'{e}, P. et al. 2006, ApJS, 165, 57
\reference{dR04} del R\'{i}o, M. S., Brinks, E., \& Cepa, J. 2004, AJ, 128, 89
\reference{Du05} Dudik, R. P., Satyapal, S., Gliozzi, M., \& Sambruna, R. M. 
2005, ApJ, 620, 113
\reference{Er02} Eracleous, M., Shields, J. C., Chartas, G., \& Moran, E. C. 
2002, ApJ, 565, 108
\reference{Er10} Eracleous, M., Hwang, J. A. \& Flohic, H. M. L. G. 2010, 
ApJ, 711, 796
\reference{Fl06} Flohic, H. M. L. G., Eracleous, M., Chartas, G., Shields, 
J. C., \& Moran, E. C. 2006, ApJ, 647, 140
\reference{Ga03} Gallo, E., Fender, R. P., \& Pooley, G. G. 2003, MNRAS, 
344, 60
\reference{Ga06} Gallo, E., Fender, R. P., Miller-Jones, J. C. A., Merloni, 
A., Jonker, P. G., Heinz, S., Maccarone, T. J., \& van der Klis, M. 2006, 
MNRAS, 370, 1351
\reference{Ga00} Garcia, M. R., Murray, S. S., Primini, F. A., Forman, W. 
R., McClintock, J. E., \& Jones, C. 2000, ApJ, 537, 23
\reference{Gr03} Graham, A. W. \& Guzm\'{a}n, R. 2003, ApJ, 125, 2936
\reference{GD04} Gonz\'{a}lez Delgado, R. M., Cid Fernandes, R.; P\'{e}rez, 
E., Martins, L. P., Storchi-Bergmann, T., Schmitt, H., Heckman, T., \& 
Leitherer, C. 2004, ApJ, 605, 127
\reference{GM06} Gonz\'{a}lez-Mart\'{i}n, O., Masegosa, J., M\'{a}rquez, I., 
Guerrero, M. A., \& Dultzin-Hacyan, D. 2006, A\&A, 460, 45
\reference{GM06} Gonz\'{a}lez-Mart\'{i}n, O., Masegosa, J., M\'{a}rquez, I., 
Guainazzi, M., \& Jim\'{e}nez-Bail\'{o}n, E. 2009, A\&A, 506, 1107
\reference{Gr07} Greene, J. E. \& Ho, L. C. 2007, ApJ, 670, 92
\reference{Ho97} Ho, L. C., Filippenko, A. V., \& Sargent, W. L. W., 1997, 
ApJS, 112, 315
\reference{Ho01} Ho, L. C. 2001, ApJ, 549, 5
\reference{Ho03} Ho, L.C., Terashima, Y., \& Ulvestad, J. S. 2003, 589, 783
\reference{Ho08} Ho, L. C. 2008, ARA\&A, 46, 475
\reference{Ka05} Kalberla, P. M. W. et al. 2005, A\&A, 440, 775
\reference{Ka04} Karachentsev, I. D., Karachentseva, V. E., Huchtmeier, W. 
K., \& Makarov, D. I. 2004, AJ, 127, 2031
\reference{Ko95} Koratkar, A., Deustua, S. E., Heckman, T., Filippenko, A. 
V., Ho, L. C., \& Rao, M. 1995, ApJ, 440, 132
\reference{Li00} Lira, P., Lawrence, A., \& Johnson, R. A. 2000, MNRAS, 319, 17
\reference{Ma98} Maoz, D., Koratkar, A., Shields, J. C., Ho, L. C., 
Filippenko, A. V., \& Sternberg, A. 1998, AJ, 116, 55
\reference{Ma05} Maoz, D., Nagar, N. M., Falcke, H., \& Wilson, A. S. 2005, 
ApJ, 625, 699
\reference{Me95} Meurer, G. R., Heckman, T. M., Leitherer, C., Kinney, A., 
Robert, C., \& Garnett, D. R. 1995, AJ, 110, 2665
\reference{MM09} Mu\~{n}oz Mar\'{i}n, V. M., Gonz\'{a}lez Delgado, R. M., 
Schmitt, H. R., Cid Fernandes, R., P\'{e}rez, E. 2009, Ap\&SS, 324, 253
\reference{Na05} Nagar, N. M., Falcke, H., \& Wilson, A. S. 2005, A\&A, 435, 
521
\reference{Ol07} Olsson, E., Aalto, S., Thomasson, M., Beswick, R., \& 
H\"{u}ttemeister, S. 2007, A\&A, 473, 389
\reference{Pl98} Plana, H., Boulesteix, J., Amram, Ph., Carignan, C., \& 
Mendes de Oliveira, C. 1998, A\&AS, 128, 75
\reference{Pe05} Peterson, B. M. et al. 2005, ApJ, 632, 799
\reference{Po00} Pogge, R. W., Maoz, D., Ho, L. C., \& Eracleous, M. 2000, 
ApJ, 532, 323
\reference{Ra01} Ravindranath, S., Ho, L. C., Peng, C. Y., Filippenko, A. 
V., \& Sargent, W. L. W. 
2001, AJ, 122, 653
\reference{Sa04} Satyapal, S., Sambruna, R. M., \& Dudik, R. P. 2004, A\&A, 
414, 825
\reference{Sc98} Schlegel, D. J., Finkbeiner, D. P., \& Davis, M. 1998, ApJ, 
500, 525
\reference{Se10} Seth, A.C. et al. 2010, ApJ, 714, 713
\reference{Sh06} Shemmer, O., Brandt, W. N., Netzer, H., Maiolino, R., 
Kaspi, S. 2006, ApJ, 646, 29
\reference{Sm01} Smith, R. K., Brickhouse, N. S., Liedahl, D. A., \& Raymond, 
J. C. 2001, ApJ, 556, 91
\reference{Te00} Terashima, Y., Ho, L. C., Ptak, A. F., Mushotzky, R. F., 
Serlemitsos, P. J., Yaqoob, T., Kunieda, H. 2000, ApJ, 533, 729
\reference{Tr01} Tremonti, C. A., Calzetti, D., Leitherer, C., \& Heckman, T. 
M. 2001, ApJ, 555, 322
\reference{Tr04} Tremonti, C. A. et al. 2004, ApJ, 613, 898
\reference{vP81} van Paradijs, J. 1981, A\&A, 103, 140
\reference{Wa91} Wang, Q., \& Helfand, D. J. 1991, ApJ, 370, 541
\reference{Wa04} Wang, J.-M., Watarai, K.-Y. \& Mineshige, S. 2004, ApJ, 607, 
107
\reference{Wu08} Wu, Q. \& Gu, M. 2008, ApJ, 682, 212
\reference{Zh09} Zhang, W. M., Soria, R., Zhang, S. N., Swartz, D. A., \& 
Liu, J. F. 2009, ApJ, 699, 281
\end{references}
\end{document}